\documentclass{ws-procs9x6}

\def\etal {{\it et al.}}

\begin{document}

\title{A NEW LORENTZ-VIOLATING MODEL OF NEUTRINO OSCILLATIONS}

\author{KEVIN R.\ LABE$^*$}

\address{Department of Physics and Astronomy, Swarthmore College,\\
Swarthmore, PA 19081\\
$^*$E-mail: klabe1@swarthmore.edu}

\begin{abstract}
A new model for neutrino oscillations is introduced, in which mass-like behavior is seen at high energies, but various behavior can be predicted at low energies.  The model employs no neutrino masses, but instead relies on the Lorentz-violating parameters $a$ and $c$.  Oscillations into sterile neutrinos and into antineutrinos are also considered.
\end{abstract}

\bodymatter

\section{Introduction}
Neutrino oscillations have been experimentally observed in a variety of situations, and are among the first evidence of physics beyond the Standard Model.  Typically, these oscillations are explained by attributing mass to neutrinos; however, not all experiments can be explained using the same masses - notably, LSND\cite{LSND} and MiniBooNE\cite{MiniBooNE2} require a larger mass-squared difference than the other experiments, and cannot be explained using a three-flavor theory of mass.  Furthermore, recent results at MINOS and MiniBooNE have hinted at an asymmetry between neutrinos and antineutrinos \cite{MiniBooNE2, MINOS}, which would be evidence for Lorentz violation.

It has already been shown that models incorporating Lorentz violations can reproduce many of the results of the mass model\cite{LongNP}.  Examples include the Bicycle Model\cite{Bicycle} and the Tandem Model\cite{Tandem}.  Here, a new model is introduced to attempt to explain these experiments.

\section{Theory}
We consider three generations of light, left-handed neutrinos, and three generations of light, sterile, right-handed neutrinos, and their antiparticles.  We allow for small mass and small general Lorentz violations.  To first order, the general Hamiltonian is a 12 $\times$ 12 matrix, given in block form by
\begin{equation}
H = \frac{1}{E} \left[ \begin{array}{cc}
H_{11} & H_{12} \\
H_{21} & H_{22} \\
\end{array} \right]
\end{equation}
where
\begin{eqnarray}
H_{11} & = & \left[ \begin{array}{cc}
-c_L^{\mu \nu}p_\mu p_\nu + a_L^\mu p_\mu & -c_M^{\mu \nu}p_\mu p_\nu + a_M^\mu p_\mu \\
-c_M^{\dagger \mu \nu}p_\mu p_\nu + a_M^{\dagger \mu}p_\mu & -c_R^{T\mu \nu}p_\mu p_\nu - a_R^{T \mu} p_\mu \\
\end{array} \right] \\
H_{12} & = & \left[ \begin{array}{cc}
-ig^{\lambda \mu \nu}_D p_\lambda q_\mu p_\nu + iH_D^{\lambda \mu}p_\lambda q_\mu & -ig_M^{\lambda \mu \nu}p_\lambda q_\mu p_\nu + iH_M^{\lambda \mu}p_\lambda q_\mu\\
-ig_M^{\dagger \lambda \mu \nu}p_\lambda q_\mu p_\nu + iH_M^{\dagger \lambda \mu}p_\lambda q_\mu & -ig_D^{T \lambda \mu \nu}p_\lambda q_\mu p_\nu - iH_D^{T \lambda \mu}p_\lambda q_\mu\\
\end{array} \right] \\
H_{21} & = & \left[ \begin{array}{cc}
ig_D^{\lambda \mu \nu}p_\lambda q^*_\mu p_\nu - iH_D^{\lambda \mu}p_\lambda q^*_\mu & ig_M^{\lambda \mu \nu}p_\lambda q^*_\mu p_\nu - iH_M^{\lambda \mu}p_\lambda q^*_\mu \\
ig_M^{\dagger \lambda \mu \nu}p_\lambda q^*_\mu p_\nu - iH_M^{\dagger \lambda \mu}p_\lambda q^*_\mu & ig_D^{T\lambda \mu \nu}p_\lambda q^*_\mu p_\nu + iH_D^{T \lambda \mu}p_\lambda q^*_\mu \\
\end{array} \right] \\
H_{22} & = & \left[ \begin{array}{cc}
-c_R^{\mu \nu}p_\mu p_\nu + a_R^\mu p_\mu & -c_M^{T\mu\nu}p_\mu p_\nu - a_M^{T\mu}p_\mu \\
-c_M^{*\mu \nu}p_\mu p_\nu - a^{*\mu}_M p_\mu & -c_L^{T \mu \nu}p_\mu p_\nu - a_L^{T\mu}p_\mu \\
\end{array} \right]
\end{eqnarray}
in the basis $\{\nu_L,\bar{\nu}_R,\nu_R,\bar{\nu}_L\}$.  Note that mass does not appear because it enters only at second order.  Such a Hamiltonian allows many unusual features, including neutrino-antineutrino mixing, neutrino-sterile neutrino mixing, strange energy dependence and direction dependence.

\section{Model}
We propose a Tricycle Model, in which we assume that $g = H = 0$ so that the off-diagonal terms can be ignored.  This allows us to restrict our attention to one quadrant of the matrix, the $\{\nu_L,\bar{\nu}_R\}$ sector ($H_{11}$).  Note that we have only considered the isotropic part of each of the Lorentz-violating coefficients.

In particular, the model to be investigated has the form
\begin{equation}
H_T = \left[
\begin{array}{cc}
0 & A \\
A^\dagger & CE \\
\end{array} \right]
\end{equation}
where $A$ is taken to be Hermitian and $A$ and $C$ commute, so that they can be simultaneously diagonalized.  We assume that the diagonalizing matrix has the conventional form\cite{PDG}
\begin{equation}
U = \left[ \begin{array}{ccc}
\cos \theta_{12} & \sin \theta_{12} & 0 \\
-\sin \theta_{12} \cos \theta_{23} & \cos \theta_{12} \cos \theta_{23} & \sin \theta_{23} \\
\sin \theta_{12} \sin \theta_{23} & -\cos \theta_{12} \sin \theta_{23} & \cos \theta_{23} \\
\end{array} \right]
\end{equation}
(we are assuming that $\theta_{13} = 0$).  The model is then fixed by 8 parameters: the two mixing angles and the three eigenvalues of each block, which we call $\{a_i\}$ and $\{c_i\}$ respectively.  The eigenvalues of the Hamiltonian are
\begin{equation}
\lambda_{i\pm} = \frac{c_iE \pm \sqrt{(c_iE)^2 + 4a_i^2}}{2}
\end{equation}

The model employs a seesaw mechanism to produce very different behavior at high and low energies.  At high energy the $CE$ matrix dominates, which cuts off left-right oscillations and allows the left-handed neutrinos to oscillate among themselves as normal.  At low energies, however, the $A$ terms dominate, and oscillations into sterile neutrinos are predicted.  Observe that
\begin{equation}
\lim_{c_iE\rightarrow \infty} \lambda_{i-} = \frac{-a_i^2}{c_iE}
\end{equation}
so that three of the eigenvalues have the expected $E^{-1}$ energy dependence at high energies.

Transition probabilities can be calculated exactly.  For example,
\begin{align}
P_{e\mu} = 4&\sin^2\theta_{12} \cos^2\theta_{12} \cos^2\theta_{23} \text{ } \times \nonumber \\ 
&[-\sin^2\alpha \cos^2\alpha \sin^2(\frac{\Delta \lambda_{41}L}{2}) - \sin^2\beta \cos^2\beta \sin^2(\frac{\Delta \lambda_{52}L}{2}) \nonumber \\
&+ \sin^2\alpha \sin^2\beta \sin^2(\frac{\Delta \lambda_{21}L}{2}) + \sin^2\alpha \cos^2\beta \sin^2(\frac{\Delta \lambda_{51}L}{2}) \nonumber \\
&+ \cos^2\alpha \sin^2\beta \sin^2(\frac{\Delta \lambda_{24}L}{2}) + \cos^2\alpha \cos^2\beta \sin^2(\frac{\Delta \lambda_{54}L}{2})]
\end{align}
where
\begin{equation}
\sin^2 \alpha = \frac{1}{2}[1 - \frac{1}{\sqrt{1+(\frac{2a_1}{c_1E})^2}}]
\end{equation}
and
\begin{equation}
\sin^2\beta = \frac{1}{2}[1 - \frac{1}{\sqrt{1+(\frac{2a_2}{c_2E})^2}}]
\end{equation}

The mixing angles and two of the eigenvalues are determined by the high-energy, mass-like behavior widely detected.  There remain four independent parameters in the model, which can be adjusted to control the low energy behavior without disrupting the high-energy limit.  As energy decreases, the probabilities diverge smoothly from the standard mass predictions.

\section{CPT violation}
The model introduced above (6) will never produce observable CPT violations.  This is because $A$ is a CPT-odd variable, but $C$ is CPT-even, so that under CPT transformations, $H$ goes to
\begin{equation}
H' = \left[ \begin{array}{cc}
0 & -A \\
-A^\dagger & CE \\
\end{array} \right]
\end{equation}
However, the eigenvalues and mixing angles on which the probability depends do not observe the sign of $A$, as can be seen from their definitions (8, 11, 12).  This causes the probabilities to be the same whether $A$ or $-A$ is used.  In fact, even if $A$ does not commute with $C$, CPT symmetry will still be preserved; to introduce CPT violation, $A$ and $C$ terms must be mixed (for example, an ordinary rotation of (6) will introduce CPT violations).

\section{Conclusion}
This model is intended to show that behavior typical of mass models can be reproduced by a Lorentz-violating model without mass.  A variety of low-energy behavior is consistent with the same behavior at high energy.  However, it remains difficult to explain all experiments, even with four free parameters.

\section*{Acknowledgments}
We would like to thank Swarthmore College for funding the research reported here.  Thanks also to Matt Mewes for his advice and assistance in this project.


\begin{thebibliography}{xx}
\bibitem{LSND}
LSND Collaboration, A.\ Aguilar \etal, Phys.\ Rev.\ D {\bf 64}, 112007 (2001).
\bibitem{MiniBooNE2}
MiniBooNE Collaboration, A.\ A.\ Aguilar-Arevalo \etal, Phys.\ Rev.\ Lett.\ {\bf 103}, 111801 (2009).
\bibitem{MINOS}
P. Vahle \etal (MINOS Collaboration), Neutrino 2010 plenary talk, June 14, 2010; to be published in the Proceedings.
\bibitem{LongNP}
V. Alan Kosteleck\'y and Matthew Mewes, Phys.\ Rev.\ D {\bf 69}, 016005 (2004).
\bibitem{Bicycle}
V. Alan Kosteleck\'y and Matthew Mewes, Phys.\ Rev.\ D {\bf 70}, 031902 (2004).
\bibitem{Tandem}
Teppei Katori, V. Alan Kosteleck\'y and Rex Tayloe, Phys.\ Rev.\ D {\bf 74}, 105009 (2006).
\bibitem{PDG}
C. Amsler \etal, Phys.\ Lett.\ {\bf B667}, 1 (2008).
\end{thebibliography}
\end{document}